\documentclass[twocolumn,floatfix]{aastex631}
\usepackage{multirow}
\usepackage[latin9]{inputenc}
\usepackage{hyperref}
\usepackage{scrextend}
\usepackage{placeins}
\usepackage{booktabs}
\usepackage{amsmath}
\graphicspath{{./}{figures/}}
\newcommand{\fsky}{f_{\rm sky}}

\begin{document}

\title{An Independent Measure of the Kinematic Dipole from SDSS}
\author[0000-0001-7888-4270]{Prabhakar Tiwari}
\affil{Department of Physics, Guangdong Technion - Israel Institute of Technology, Shantou, Guangdong 515063, P.R. China}
\author[0000-0003-2413-0881]{Dominik J. Schwarz} 
\affil{Fakult{\"a}t f{\"u}r Physik, Universit{\"a}t Bielefeld, Postfach 100131, 33501 Bielefeld, Germany}
\author[0000-0003-4726-6714]{Gong-Bo Zhao}
\affiliation{National Astronomical Observatories,
Chinese Academy of Sciences, Beijing, 100101, P.R.China}
\affiliation{School of Astronomy and Space Science, University of Chinese Academy of Sciences, Beijing, 100049, P.R.China}
\affiliation{Institute for Frontiers in Astronomy and Astrophysics, Beijing Normal University, Beijing, 102206,P.R.China}
\affiliation{Chinese Academy of Sciences South America Center for Astronomy (CASSACA), National Astronomical Observatories of China, Beijing, 100101, P.R.China}
\author[0000-0001-9833-2086]{Ruth Durrer}
\affiliation{D\'epartement de Physique Th\'eorique, Universit\' de Gen\`eve,
Quai E. Ansermet 24. 1211 Gen\`eve, Switzerland}
\author[0000-0002-3052-7394]{Martin Kunz}
\affiliation{D\'epartement de Physique Th\'eorique, Universit\' de Gen\`eve,
Quai E. Ansermet 24. 1211 Gen\`eve, Switzerland}
\author[0000-0002-8800-5740]{Hamsa Padmanabhan}
\affiliation{D\'epartement de Physique Th\'eorique, Universit\' de Gen\`eve,
Quai E. Ansermet 24. 1211 Gen\`eve, Switzerland}

\begin{abstract}

We utilize the Sloan Digital Sky Survey (SDSS) extended Baryon Oscillation Spectroscopic Survey (eBOSS) and Baryon Oscillation Spectroscopic Survey (BOSS) catalogs with precise spectroscopic redshifts to estimate the kinematic redshift dipole caused by the proper motion of the Solar system. We find that the velocity extracted from the kinematic dipole is consistent with Cosmic Microwave Background inferred values. Although the small sky coverage and limited number density of the SDSS sources constrain us from obtaining precise and robust measurements, we leverage the redshift dipole method to estimate the kinematic dipole. The velocity measurements in this study are insensitive to intrinsic clustering, associated with the source count dipole. The kinematic dipole measured in this work and its consistency with CMB values do not guarantee isotropy at large scales. The anisotropy (excess dipole) measured with the NRAO VLA Sky Survey (NVSS) and the WISE Catalog (CatWISE) could be due to the intrinsic distribution of galaxies. The results in this work focus solely on the kinematic dipole term.

\end{abstract}
\keywords{Cosmology (343); Cosmological principle (2363), Large-scale structure of the universe (902);  Observational cosmology (1146); Galaxy redshifts(1378)}

\section{Introduction}
\label{sc:intro}
The Cosmological Principle, one of the fundamental assumptions of the standard $\Lambda$CDM cosmological model  assumes that our Universe  is statistically homogeneous and isotropic \citep{Milne:1933CP,Milne:1935CP}. The strongest observational support for this assumption comes from the Cosmic Microwave Background (CMB), which is uniform to roughly 1 part in $10^5$ \citep{Penzias:1965,COBE:1993,COBE_White:1994,WMAP:2013,Planck_iso:2020},  apart from a dipole to which we will turn below. However,  the CMB  provides a picture of the early Universe and the observed isotropy in the CMB traces the density perturbations that existed around 370,000 years after the Big Bang, when neutral hydrogen was formed. The density perturbations in the early Universe grew, leading to the formation of galaxies, galaxy clusters, and all cosmological structures. Even a very small anisotropy present in the early Universe could have resulted in an anisotropic distribution of matter later on. Furthermore, it is not sufficient that one observer sees a nearly isotropic CMB for the Universe to be approximated by a
Friedmann-Lema\^\i tre  model. Instead, one has to require that an entire congruence of observers see a nearly isotropic CMB \citep{Stoeger:1994qs}. Therefore, the observed near isotropy of the CMB does not necessarily guarantee isotropy and homogeneity of matter distribution at late times. An independent test of the Cosmological Principle at late times, is thus needed. Conventionally this test is performed by measuring the large-scale angular clustering, i.e., the low multipoles of a large-scale structure (LSS) tracer \citep{Blake:2002,Tiwari:2019l123}. This is frequently done with radio galaxy surveys, which offer a uniform survey over a large sky area and are suitable for large-scale clustering signal analysis. In addition to galaxy clustering, other powerful probes of the Cosmological Principle include the anisotropy in radio polarization offset angles \citep{Jain:1998kf}, alignment of quasar polarizations \citep{Hutsemekers:1998}, alignment of radio galaxy axes and the correlation of their polarization vectors across varying distance scales \citep{PTiwari:2013pol,Tiwari:2016si,Tiwari:2019,Tiwari-GA:2021}, bulk flow in X-ray clusters \citep{Kashlinsky:2010,Migkas:2021}, isotropy exploration with ultra high energy cosmic rays \citep{Abu-Zayyad:2012,Takami:2016,TelescopeArray:2020}, angular distribution of Gamma-ray bursts \citep{Balazs:1999,Meszaros:2003,Tarnopolski:2017}, fast radio bursts distribution \citep{Qiang:2019}, inference of cosmic rest-frame from supernovae \citep{Horstmann:2021jjg,Sorrenti:2022zat,Sorrenti:2024}, and the cosmic bulk flows \citep{Watkins:2023rll} etc. 

Despite the very smooth and isotropic CMB, the CMB itself exhibits a dipole anisotropy at a level of 1 part in $10^3$. This is two orders of magnitude larger than the fluctuations on smaller scales. However, this dipolar anisotropy is commonly understood to be due to  the motion of the Solar system with respect to the CMB rest frame. The dipole in the CMB temperature map \citep{Conklin:1971,Henry:1971,Corey:1976,Smoot:1977,Kogut:1993,Hinshaw:2009, Planck_I:2020,Planck_II:2020,Planck_III:2020}, if entirely kinematic, this translates to a very precise value of the Solar system motion with respect to the CMB frame, corresponding to a speed of $369.82 \pm 0.11$ km s$^{-1}$ in the direction $l=264^\circ.021 \pm 0^\circ.011$, $b=48^\circ.253\pm 0^\circ.005$ in galactic coordinates \citep{Planck_dipole:2020}. 

A crucial independent measure of the Solar system net motion with respect to the CMB or a distant rest frame---such as any LSS tracer at sufficient distances where peculiar velocities average out and become small in comparison---is warranted. Within standard cosmology, the overall mass distribution and the LSS are expected to share the same rest frame as the CMB, and hence the LSS should contain a matching dipole anisotropy from our motion. For this purpose, radio source catalogs have long been utilized, primarily due to the extensive and uniform sky coverage provided by radio continuum surveys, to explore large-scale structure (LSS) clustering. In particular, radio continuum catalogs have been used to determine the dipole signal \citep{Blake:2002, Singal:2011, Gibelyou:2012, Rubart:2013, Tiwari:2014ni, Tiwari:2015np, Tiwari:2016adi, Dolfi:2019, Siewert:2020CRD, Darling:2022, Wagenveld:2023, Singal:2024, Cheng:2024, Oayda:2024} that presumably emerges in radio galaxy number counts and flux distribution largely due to Doppler and aberration effects caused by our local motion \citep{Ellis:1984, Tiwari:2014ni}. Additionally, there have been other efforts using different tracers and observables to probe this kinematic velocity and the cosmological principle \citep{Schwarz:2004,Kalus:2013,Schwarz:2016, Bengaly:2016, Secrest:2020CPQ, Secrest:2022,Horstmann:2021jjg,Dam:2022,Kothari:2024, Oayda:2023, Abghari:2024, daSilveiraFerreir:2024, Mittal:2024,Sorrenti:2024}.

It is important to note that the conventional and most widely explored method to determine our motion with respect to the CMB and to probe the cosmological principle---the dipole estimation from galaxy catalogs---involves several important concerns. The most obvious concern is the systematics in data and estimators used \citep{Tiwari:2019TGSS, Siewert:2020CRD, Guandalin:2022}. Furthermore, the measurement of the dipole includes the clustering dipole and the shot-noise dipole in addition to the kinematic dipole. The clustering dipole, assuming $\Lambda$CDM, is relatively small if LSS is concentrated at redshifts $\sim0.5$ or higher, and the shot-noise dipole, inversely proportional to the square root of the number density, also becomes insignificant if the number density is high enough. Reliable dipole signal estimation requires more than 50\% of sky coverage \footnote{The exact sky coverage required depends on the dipole signal magnitude and direction, survey footprint, and source density.} \citep{Tiwari:2014ni,Bengaly:2019}.

The kinematic dipole (both direction and magnitude) can be estimated assuming the CMB velocity, spectral index around the observed frequency, and the dependence of the total number count on the flux density cut \citep{Ellis:1984,Tiwari:2015np}. The shot-noise and clustering dipole magnitudes can also be estimated given the number density, redshift distribution of sources, and the galaxy bias for LSS populations. However, the direction of these two dipoles is random, adding uncertainty to the observed dipole from LSS tracers. The problem is mitigated by assuming these two dipoles are insignificant compared to the kinematic dipole. However, almost all studies\footnote{For a comprehensive discussion, see \citet{Leandros:2021, Peebles:2022, Aluri:2022}.} have found significantly larger dipoles than expected, indicating either the presence of serious (common) systematics or a significantly large kinematic and/or clustering dipole, thus suggesting a potential violation of the cosmological principle \citep{Aluri:2022, Secrest:2022}. Remarkably, most studies find the dipole direction close to the CMB kinematic dipole, suggesting either that the Solar system motion with respect to the CMB frame is significantly large or that the clustering dipole is significantly large and its direction aligns closely with the CMB kinematic dipole. This makes it critically important to separate the clustering and kinematic dipoles in LSS dipole estimates. The first exploration of this was proposed by suggesting an improved fit to the number count versus flux cut for NRAO VLA Sky Survey (NVSS; \citealt{Condon:1998}) sources \citep{Tiwari:2015np}. Subsequently, \citet{Singal:2019} proposed estimating the redshift dipole to obtain an independent estimate of the kinematic dipole. Later, \citet{Nadolny:2021} presented a detailed analysis to separate the kinematic and intrinsic dipoles by investigating the number counts dipole along with the dipole in the distribution of galaxy fluxes, sizes, and/or redshifts. \cite{Horstmann:2021jjg} measured the kinematic dipole from supernovae of type 1a and found it to be consistent with the CMB inferred kinematic dipole, but~\cite{Sorrenti:2022zat,Sorrenti:2024}, using a more recent data set found a disagreement with the CMB dipole direction.

Recently, \citet{daSilveiraFerreir:2024} employed the Sloan Digital Sky Survey (SDSS) extended Baryon Oscillation Spectroscopic Survey (eBOSS) and the Baryon Oscillation Spectroscopic Survey (BOSS) catalogs to explore the redshift dipole and presented their analysis on the kinematic dipole. They concluded that the velocities extracted from the kinematic dipole are consistent with those expected from the CMB kinematic dipole, arguing that their findings provide significant empirical support for the cosmological principle. 

In this paper, we revisit the SDSS eBOSS and BOSS catalogs and present our detailed analysis. We compute the intrinsic dipole for all LSS tracers in the study and find that some of the BOSS/eBOSS tracers exhibit a comparable clustering dipole following $\Lambda$CDM. We further find with mocks that the redshift dipole is almost insensitive to the clustering dipole, and thus the redshift dipole estimation does not allow any conclusions on the clustering dipole and, consequently, on the cosmological principle. However, it provides an independent measure of the kinematic dipole and hence an independent measure of our local motion with respect to distant LSS. Due to limited sky coverage (at most 26\% with SDSS tracers), determining the monopole becomes unreliable, resulting in significant uncertainty in the dipole estimation. We assume a prior value for the dipole to obtain consistent results and are only able to perform a consistency test to compare our local motion inferred from the CMB with SDSS observations.

Our analysis, approach, and conclusions differ from \citet{daSilveiraFerreir:2024}. The main difference is that we allow for a large clustering dipole, as observed in radio and infrared surveys, and demonstrate that the redshift dipole is not affected by it. Furthermore, \citet{daSilveiraFerreir:2024}'s main focus is on the cosmological principle, and they claim to explore anisotropy with the redshift dipole. However, we differ in our approach and conclusions, asserting that this is an independent measure of the kinematic dipole. There are other differences as well, such as the way they use weights and their methodology.

If our tentative results are confirmed with larger surveys like Euclid~\citep{Euclid:2024yrr}, it will allow us to infer that the matter and CMB rest frames agree, as we expect in standard cosmology. 

We discuss the data used in this work in Section \ref{sc:data}. Next, we detail the estimator used and our method in Section \ref{sc:method}. In Section \ref{sc:results}, we present our results, and we conclude with a discussion in Section \ref{sc:dc}.

\section{Data}
\label{sc:data}
We aim to utilize all viable large-scale structure (LSS) tracers from SDSS BOSS and eBOSS, focusing on datasets with well-understood data quality, minimal systematics, and substantial sky coverage. For this purpose, we utilize clustering catalogs finalized and provided by the SDSS collaboration. These catalogs include observational weightings and mocks that yield cosmological clustering results consistent with the standard $\Lambda$CDM cosmology framework. Specifically, we employ five tracers, the same as used in \cite{daSilveiraFerreir:2024}: quasars (QSOs; \citealt{Ata:2018QSO}), luminous red galaxies (LRGs; \citealt{Bautista:2018LRG}), and massive galaxies in eBOSS DR16, i.e., CMASS, clustering catalogs \citep{Ross:2020DR16} along with BOSS DR12 clustering catalogs, which include LOWZ and CMASS \citep{Reid:2016DR12}. We also employ the redshift cuts identical to \cite{daSilveiraFerreir:2024} to avoid data systematics. 

\begin{figure*}
    \centering
    \includegraphics[width=1.0\textwidth]{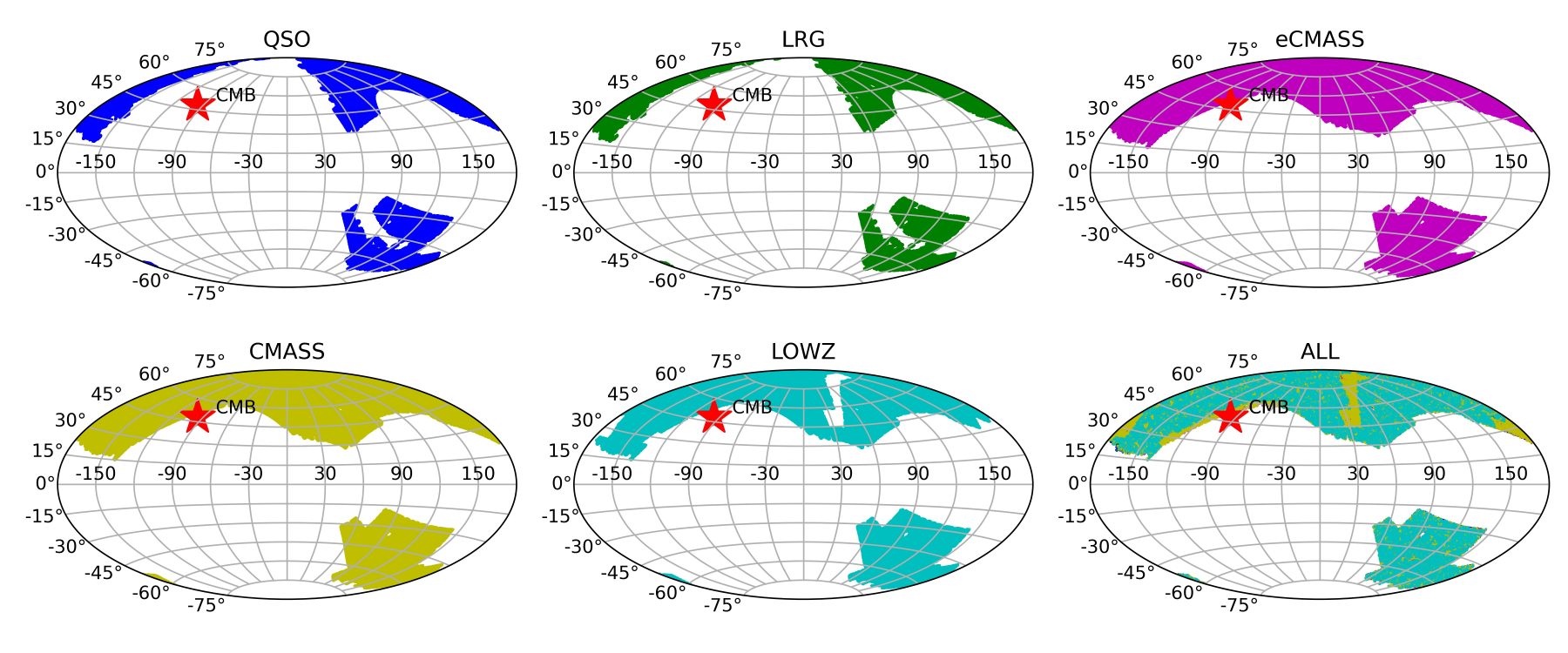}\\
    \caption{The footprints of BOSS and eBOSS tracers in galactic coordinates, along with the location of the CMB dipole (shown as a red star), are illustrated. The panel displaying eCMASS galaxies represents the CMASS galaxies from eBOSS. The final panel shows all the tracers, with only 1\% of the sources displayed for visibility.}
    \label{fig:all_tracers}
\end{figure*}

\begin{figure}[htbp]
    \centering
    \includegraphics[width=0.48\textwidth]{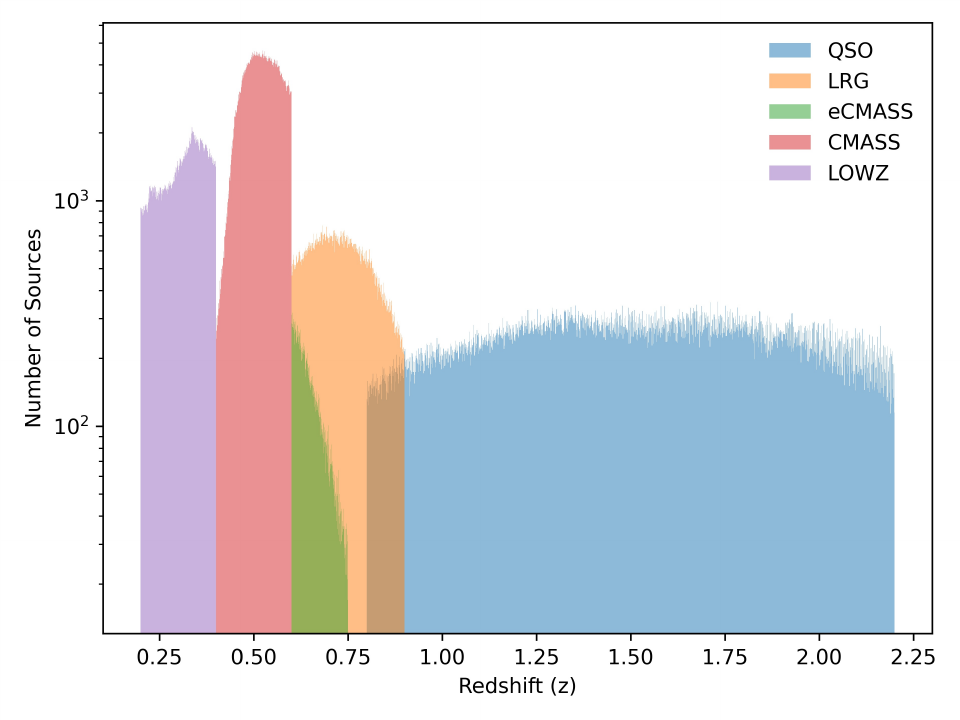}
    \caption{The redshift distribution of BOSS and eBOSS tracers used in this work.}
    \label{fig:dndz}
\end{figure}

Mock catalogs for these LSS tracers are available from the SDSS collaboration \citep{Kitaura:2016DR12mock,Zhao:2021eBOSSmock}, and we employ them in our analysis for error estimation. It is noteworthy that the mock catalogs neither include a number density nor a redshift dipole, and we modify the mocks to include both dipole anisotropy features. The sky coverage for these LSS tracers is quite small, reaching a maximum of up to 26\%, and therefore we are bound to encounter large errors in our estimates. For convenience, the redshift range and the total number of sources in each tracer are listed in Table \ref{tb:data}. Figure \ref{fig:all_tracers} illustrates the distribution of all tracers, including their positions relative to the CMB dipole location. Figure \ref{fig:dndz} shows the redshift distribution for the tracers after applying the redshift cuts. The exact links for data and mocks are detailed in Appendix \ref{sc:datalinks}.

\begin{deluxetable*}{c c c c c c}
\tablecaption{Tracers of LSS used in this work. We apply the same redshift cuts as in \cite{daSilveiraFerreir:2024} and report the sky fraction, $\fsky$,  covered by the various tracers and the total number of sources after applying the redshift cuts. \label{tb:data}}
\centering 
\tablehead{ & LSS tracer & redshift range  & $\fsky$ &  \#sources  } 
\startdata
\multirow{3}{*}{eBOSS} &  QSOs & $0.8<z<2.2$  & 0.14 & 343,708 \\
                     &  LRGs   & $0.6<z<0.9$  & 0.14 & 163,249 \\
                     &  CMASS & $0.6<z<0.75$ & 0.25 & 193,298 \\
                     &&&&\\
\multirow{2}{*}{BOSS} & CMASS & $0.4<z<0.6$  & 0.26 & 620,292 \\
                    &  LOWZ  &  $0.2<z<0.4$ & 0.26 &  280,067\\
\enddata
\end{deluxetable*}

\section{Estimator and method}
\label{sc:method}
The redshift dipole in galaxy surveys has been discussed in \cite{Maartens:2018qoa, Singal:2019, Nadolny:2021}, with some differences in terminology and methods. In this section, we provide a concise discussion on the redshift dipole, along with an overview of the methodology employed in this study, to ensure clarity and comprehensiveness. For an observer traveling with a velocity $v$ with respect to the rest frame of a distant large-scale structure (LSS) tracer, the light from a galaxy in direction $\hat{n}$, making an angle $\theta$ with velocity $v$, is Doppler-shifted to a frequency $\nu$. Following the formulation from special relativity, we have:
\begin{equation}
    \nu = \nu_o \delta(\hat{n}),
\end{equation}
where $\nu$ and $\nu_o$ are the observed and emitted frequencies respectively, and $\delta(\hat{n}) = \gamma \left(1+ \hat{n} \cdot \vec{\beta}\right)$, with $\gamma =1 /\sqrt{1-\beta^2}$, $\beta = v/c$, and $\hat{n} \cdot \vec{\beta} = \beta \cos \theta$. Additionally, there is a change in the galaxy position due to the aberration effect. The apparent direction $\hat{n}'$  deviates from $\hat{n}$ and is given by \citep{Jackson:1998nia, Yasini:2020,daSilveiraFerreir:2024},
\begin{equation}
\label{eq:aberration}
    \hat{n}' = \frac{\hat n \cdot \hat \beta + \beta}{1+ \hat n \cdot  \vec \beta}  \hat \beta + \frac{\hat n - (\hat n \cdot \hat \beta) \hat \beta}{\gamma (1+ \hat n \cdot \vec \beta)}. 
\end{equation}
The redshift $z$ is defined as the ratio of the difference between observed and emitted wavelengths over the emitted wavelength of electromagnetic radiation, $z = (\lambda - \lambda_0) / \lambda_0 = (\nu_o - \nu) / \nu$. 
Thus, the redshift due to the  peculiar velocity of the Solar system with respect to LSS, $z_{\rm kin}$, is given by:
\begin{equation}
\label{eq:zpec}
    1 + z_{\rm kin} = \delta(\hat{n})^{-1}.
\end{equation}

Subsequently, if the galaxy is at cosmological redshift $z_0$, then the observed redshift $z$ can be expressed as:
\begin{eqnarray}
\label{eq:doppler}
    1 + z &=& (1 + z_0) (1 + z_{\rm kin}) (1+ z_{\rm pec}) (1 + z_{\rm grav}) \nonumber \\
          &\approx& (1 + z_0) \delta(\hat{n})^{-1}, 
\end{eqnarray}
where $z_{\rm pec}$ is the redshift caused by the peculiar motion of the source, which is negligible at large enough cosmological redshift, and $z_{\rm grav}$ is the gravitational redshift, which is only significant in situations where a strong gravitational field is present. The general expression for $\delta(\hat n)$ to first order in perturbation theory can be found e.g. in \cite{Bonvin:2005ps}. Here we assume that both, the source velocity and the gravitational field are sub-dominant and the main contribution to the dipole  comes from the observer velocity, like in the CMB. We can further simplify equation (\ref{eq:doppler}) by assuming the non-relativistic limit, i.e., $\beta=v/c \ll 1$. Indeed, this approximation is supported by results from CMB and NVSS dipole measurements, indicating $v \ll c$. We have to first order in $\beta$
\begin{eqnarray}
\label{eq:z_dipole}
1 + z & \approx & (1 + z_0)(1 - \beta \cos \theta) \nonumber \\
     z & \approx & z_0 - (1+z_0) \beta \cos \theta
\end{eqnarray}
which shows that the observed redshift will possess a dipole asymmetry due to our motion with respect to the LSS, with its cosmological redshift $z_0$ as the monopole term and the dipole magnitude\footnote{Note the minus sign! The redshift dipole is in the opposite direction to the frequency dipole and, consequently, to the number density and flux density dipoles \citep{Ellis:1984,Tiwari:2014ni,Tiwari:2015np}.} as $(1+z_0) \beta/z_0$. We add that for all calculations in this work, we have utilized the relativistic form of $\delta(\hat{n}) = \gamma \left(1+ \hat{n} \cdot \vec{\beta}\right)$, and the approximation above is only to demonstrate the redshift dipole term due to a moving observer.
For a uniform sample over all the sky, the dipole term averages out, and by taking the average over the redshifts, we can determine $z_0$, i.e., $z_0 = \bar{z}$. However, for a partial sky sample such as we have with SDSS, we cannot simply take the average over observed redshifts to get the monopole term. Instead, if we assume $\vec{\beta}$ a priori, we can compute the deboosted redshifts by inverting equation (\ref{eq:doppler}), thus obtaining $z_0$. The SDSS LSS surveys have very precise redshift measurements; however, they do not encompass a large sky coverage. As listed in Table \ref{tb:data}, they cover at most 26\% of the sky. Thus, we can only perform a consistency test with SDSS tracers to determine whether our local motion extracted from the CMB is consistent with the available SDSS observations.

To obtain our local motion, $\vec \beta$, from SDSS tracers, we closely follow the method outlined in \cite{daSilveiraFerreir:2024}.  We perform a tomographic least-squares fit of the Doppler modulation around the de-Dopplered redshift monopole, using redshift bins of $\Delta z = 0.001$, which is the typical error we expect in redshift measurements from SDSS spectra \citep{Dawson:2013}. This is also approximately the error in the redshift values provided in the SDSS clustering catalogs analyzed in this paper. Since the monopole for the north and south galactic caps differs significantly \citep{daSilveiraFerreir:2024}, we compute the monopole separately for the North Galactic Cap (NGC) and the South Galactic Cap (SGC). For a given $\vec \beta$, the de-Dopplered monopole of a redshift bin is obtained by inverting equation (\ref{eq:doppler}) and averaging over all sources in the bin,
\begin{equation}
\label{eq:monopole}
Z_{0,\rm bin} (\vec{\beta}) = \frac{\sum\limits_i w_i \left[(1+z_i)\delta(\hat{n}_i) - 1\right]}{\sum\limits_i w_i},
\end{equation}
where the sum runs over all objects in the redshift bin, and $w_i$ represents the weights for each object. For eBOSS and BOSS, these weights are,
\begin{eqnarray}
\label{eq:weights}
w_{i, \rm eBOSS} & = & w_{i,\rm sys}  w_{i,\rm noz}  w_{i,\rm cp} \nonumber, \\
w_{i, \rm BOSS} & = & w_{i,\rm sys}  (w_{i,\rm noz} + w_{i,\rm cp} - 1), 
\end{eqnarray}
where $w_{i, \rm sys}$ represents the total systematic weight, $w_{i, \rm noz}$ is the redshift failure weight, and $w_{i, \rm cp}$ is the close pairs weight, which accounts for galaxies not allocated a fiber due to fiber collisions \citep{Anderson:2012}.

Subsequently, we calculate the $\chi^2$ for the redshift bin,
\begin{equation}
\chi^2_{\rm bin} (\vec{\beta}) = \frac{\sum\limits_i w_i^2 \left[z_i - \left([1+Z_{0,\rm bin} (\vec{\beta} )]\delta(\hat{n}_i)^{-1} - 1\right)\right]^2}{\sum\limits_i w_i^2} .
\end{equation}

We compute the best-fit dipole for a given LSS by minimizing the total $\chi^2$ over all bins, i.e., by minimizing, 

\begin{equation}
\label{eq:chitotal}
\chi^2 (\vec{\beta}) = \sum\limits_{\rm bin} \chi^2_{\rm bin, NGC} (\vec{\beta}) + \sum\limits_{\rm bin} \chi^2_{\rm bin, SGC} (\vec{\beta})\,
\end{equation}

for $\vec{\beta}$. Note that while fitting the dipole, we ignore the aberration effect, which slightly changes the angular positions of sources, as it is insignificant given the small value of $\vec{\beta}$. Additionally, the redshift dipole is derived solely from the redshift values of galaxies and does not directly correlate with the spatial clustering caused by LSS. In thin tomographic slices of redshift, the contribution of the clustering dipole remains insignificant, allowing for a more accurate estimation of the redshift dipole, which is fundamentally a kinematic effect caused by the Doppler shift due to our peculiar velocity. This kinematic dipole, therefore, provides a unique and direct measurement of our velocity vector with respect to the cosmic rest frame, independent of the complexities of the underlying matter distribution and its anisotropies. We will explicitly verify this with mocks in the following sections of the paper.

\subsection{Error and Bias Estimations}
\label{ssc:error}

The error and bias in our estimation are explored using mocks provided by the SDSS collaboration \citep{Kitaura:2016DR12mock,Zhao:2021eBOSSmock}. As these mocks do not contain a dipole, we first modify the redshifts by adding the Doppler boost, assuming the CMB kinematic dipole. This modification follows equations \eqref{eq:aberration} and \eqref{eq:doppler}.  Additionally, we introduce a Gaussian error with variance $\delta z = 0.001$ to account for uncertainties in measurements. The eBOSS EZmocks provide weights identical to the data, so we utilize equation \eqref{eq:weights} to calculate the total weight. For BOSS, however, we utilize MultiDark-Patchy mocks \citep{Kitaura:2016DR12mock,Rodriguez-Torres:2016DR12mocks}, which use slightly different terminology for weights. As described on their website\footnote{\href{https://www.skiesanduniverses.org/page/page-3/page-15/page-9/}{https://www.skiesanduniverses.org/page/page-3/page-15/page-9/}}, we employ $w_i=(C7 \times C8)/(1+10000 \times C5)$ to get total weight, where C5, C7, and C8 represent the number density, the veto flag, and the fiber collision weight, respectively. 

\begin{figure}[htbp]
    \centering
    \includegraphics[width=0.5\textwidth]{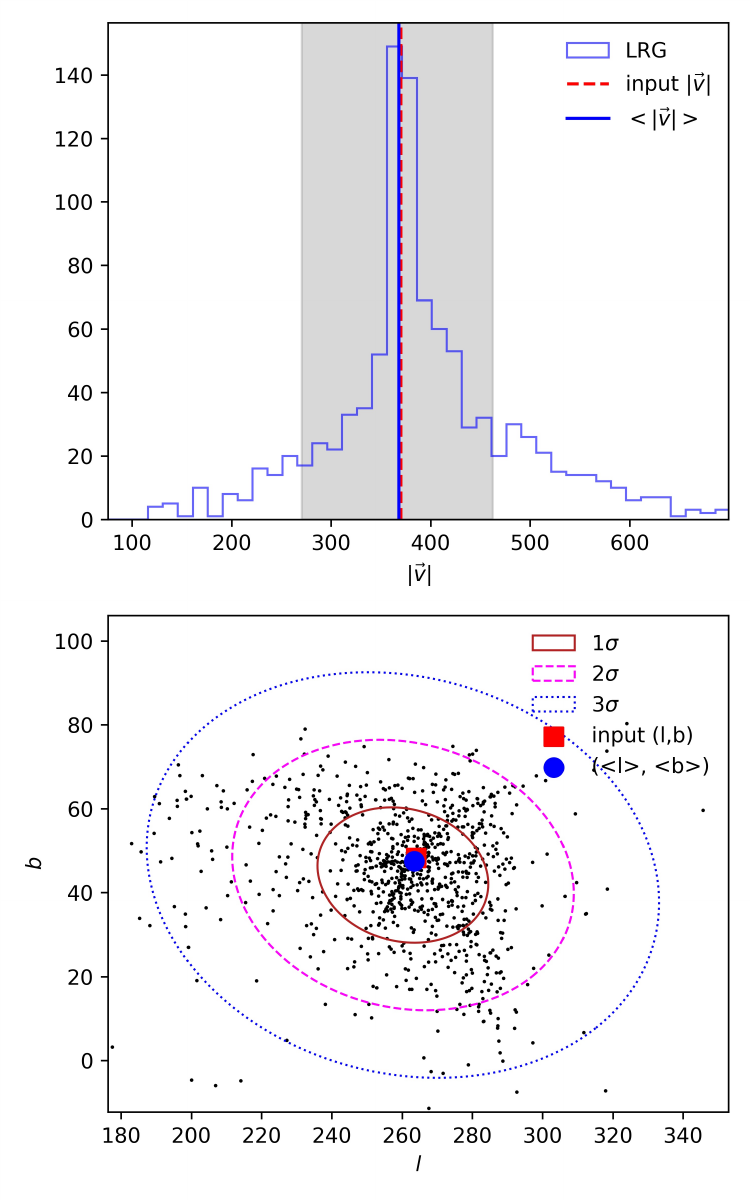}
    \caption{The dipole recovery using 1000 EZmocks for LRG tracers. The top panel shows the histogram of the recovered velocity magnitude, and the bottom panel displays the recovered angular position relative to the input (red square) direction. The large blue dot indicates the mean angular position from mocks. The 1, 2 and $3\sigma$ contours indicate  measurement errors.}
    \label{fig:EZmocks_LRG}
\end{figure}

Subsequently, we apply the same pipeline discussed above to recover the input dipole signal. With each of the mocks, we have \( v_x \), \( v_y \), and \( v_z \). To determine the mean values and error bars for \( v \), \( l \), and \( b \), we first calculate the mean and variances of \( v_x \), \( v_y \), and \( v_z \). We then convert the mean Cartesian components to spherical coordinates to obtain the mean values of \( v \), \( l \), and \( b \). The error bars are derived by propagating the uncertainties from the Cartesian components to spherical coordinates. We present the results from EZmocks for LRG tracers in Figure \ref{fig:EZmocks_LRG}. The top panel shows the recovered velocity amplitude from 1000 EZmocks. The bottom panel demonstrates our ability to recover the dipole direction. The blue dot is the mean from all mocks. The expected error in the mean is $\sigma/\sqrt{1000}$.  

The mean values of the recovered velocity and direction from mocks differ slightly from the input; this difference, termed our recovery bias, depends on the relative positions of the mask and the CMB dipole direction. However, since this bias is small compared to the scatter (error bars), and we account for it by adjusting the dipole by the same amount observed in the mocks, our results are unlikely to be affected by this bias. The error bars, determined using the 16th and 84th percentiles, are represented by the shaded gray region in the top panel of Figure \ref{fig:EZmocks_LRG}. 

\begin{figure}[htbp]
    \centering
    \includegraphics[width=0.5\textwidth]{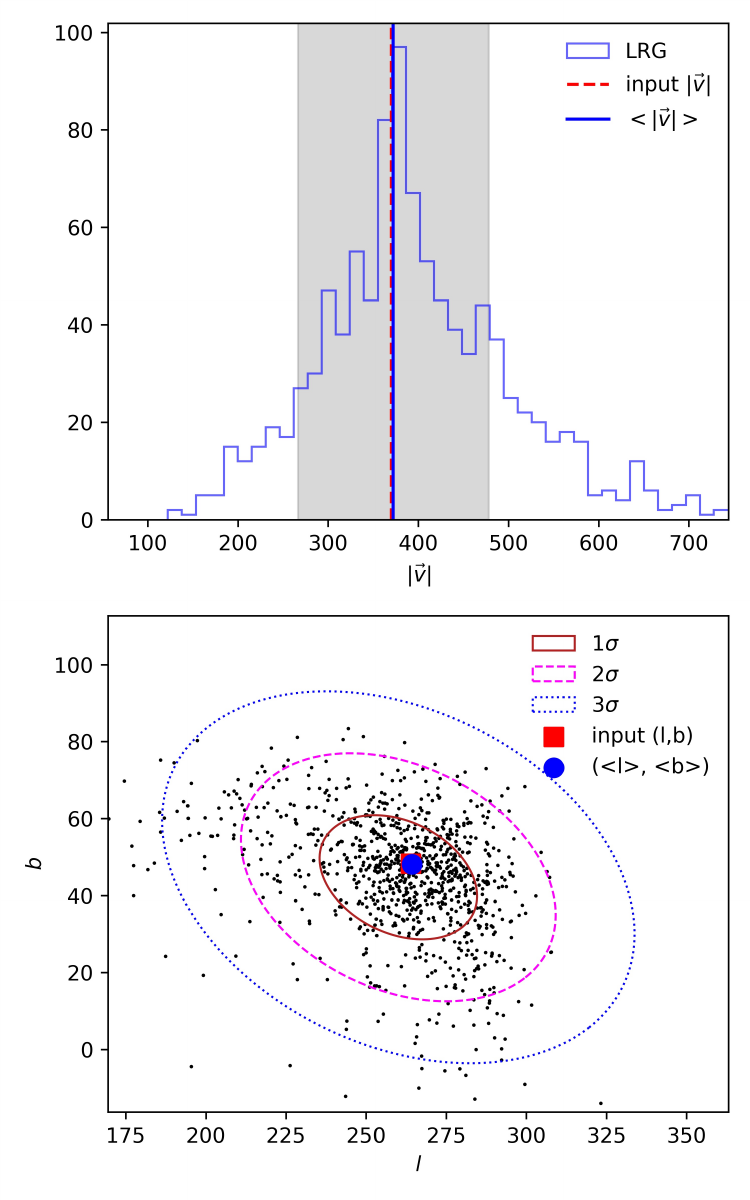}
    \caption{The dipole recovery using 1000 random mocks which includes a clustering dipole of $D = 0.86 \times 10^{-2}$ in the direction $(l, b) = (217^\circ, +20^\circ)$. Other details are the same as in Figure \ref{fig:EZmocks_LRG}.}
    \label{fig:NDD_LRG}
\end{figure}

Next, we generate random number density maps with a clustering dipole signal, following the survey's redshift \( dN/dz \) and survey footprint masks, to demonstrate the effect of the clustering dipole on the redshift dipole. Specifically, we consider a clustering dipole of $D = 0.86 \times 10^{-2}$ in the direction $(l, b) = (217^\circ, +20^\circ)$, identified as the residual dipole after subtracting the kinematic dipole in NVSS and WISE samples \citep{Secrest:2022} \footnote{In mocks, the intrinsic dipole and clustering dipole are essentially the same, both referring to galaxy clustering. However, in the literature, ``clustering dipole" usually denotes the clustering expected from the $\Lambda$CDM, while ``intrinsic dipole" refers to any residuals after accounting for this expected clustering.}. Except, for the LOWZ dataset, this value is higher than expected under $\Lambda$CDM cosmology, see~\cite{Nadolny:2021}, Fig.~1. Note also that the expected shot noise dipole, given by $D^{\mathrm{(SN)}}=3\sqrt{f_{\rm sky}/N}$ which contributes to the clustering dipole is of the order of $(2$ -- $4)\times 10^{-3}$. Despite its large amplitude, we find that the clustering dipole has almost no effect on the redshift dipole, making this an independent test of the kinematic dipole. Our random number density mocks contain only a dipole signal and not higher-order clustering multipoles as per the $\Lambda$CDM power spectrum. However, since the clustering dipoles considered in this work dominate compared to other multipoles, the results are largely unaffected by the exclusion of higher multipoles. We use these mocks to demonstrate that even a significant clustering dipole does not affect the recovery of our local motion using the redshift dipole. The dipole recovery with the clustering dipole for the LRG sample is shown in Figure \ref{fig:NDD_LRG}. It is evident that the inclusion of the clustering dipole does not affect the recovery of our local motion with respect to the LSS. We obtain similar results with mocks for all other tracers studied in this work, including the BOSS MultiDark-Patchy mocks, which show comparable recovery patterns.

Based on the previously described mocks, we use the CMB dipole as a fiducial model for 
estimating bias and error bars, assuming that the observed CMB dipole is a reliable reference and should be considered a strong prior. However, this approach does introduce model bias into our estimates. To mitigate this bias, we employ bootstrap sampling, which derives error bars directly from the data. This method ensures robust error estimation by accounting for sampling variability and providing a reliable measure of uncertainties  without relying on a  fiducial dipole.  Specifically, we generate 1000 bootstrap samples from the data by resampling with replacement. For each sample, we calculate \( v_x \), \( v_y \), and \( v_z \), and then compute the mean values and error bars for \( v \), \( l \), and \( b \) as previously discussed. Since the bootstrap samples do not include a fiducial dipole it is meaningless to correct for any potential recovery bias in this case.

\section{Results}
\label{sc:results}
We employ the method discussed in Section \ref{sc:method} and perform the least square fit to obtain the best fit $\vec \beta$ values and obtain our motion with respect to SDSS LSS tracers. In particular, we utilize the python \texttt{scipy.optimize.minimize} function to perform the optimization of our model parameters by minimizing the chi-squared statistic, which measures the goodness of fit between our binned data and the theoretical model \citep{Virtanen:2020SciPy}. Specifically, we use the Limited-memory Broyden-Fletcher-Goldfarb-Shanno algorithm for Bound-constrained (L-BFGS-B) optimization. This method is chosen due to its efficiency in handling large-scale problems and its ability to manage bound constraints directly. Notably, we fit \(\beta_x\), \(\beta_y\), and \(\beta_z\) to the data, considering bounds on these values in the range \([-0.03, 0.03]\). This range for \(\vec{\beta}\) is significantly higher than the CMB inferred velocity, allowing us to explore a broader parameter space. We input the CMB inferred kinematic dipole, i.e., $v =369$ km s$^{-1}$ in the direction $l=264^\circ$, $b=48^\circ$ as the starting point for the optimization algorithm. After correcting for recovery bias and estimating error bars from EZmocks and MultiDark-Patchy mocks as described in Section \ref{ssc:error}, we present the results for all tracers in the Table \ref{tb:results1} under the heading ``CMB dipole as initial guess". Notably, the velocity values and dipole directions recovered from all SDSS tracers except LOWZ are consistent with the CMB-observed velocity values. This largely suggests consistency with the CMB dipole, indicating that our local motion with respect to distant LSS tracers is the same as with the CMB. This affirms that the dipole in the CMB is indeed kinematic, and the velocities extracted are apparently consistent with LSS as well.  

However, while performing the $\chi^2$ minimization, we assume the CMB kinematic dipole as a starting point, which significantly affects the recovery process. To ensure robustness and independence from initial guess, we utilize \texttt{scipy.optimize.differential\_evolution}  for $\chi^2$ minimization. This method comprehensively explores parameter space without relying on any particular starting point, thereby providing unbiased best-fit dipole values. In contrast to `L-BFGS-B', which is a local optimizer, `differential\_evolution' is a global optimizer that can escape local minima and explore the search space more broadly. However, using this method, the recovery shows inconsistency with large scatter, indicating significant dependence on the choice of initial guess. The results are tabulated in Table \ref{tb:results1} under the heading ``No initial guess" and, in general, are consistent with both zero and the CMB-inferred kinematic dipole. Choosing an initial guess away from the CMB direction indeed changes the best fit results. This reflects the fact that, given the limited sky coverage of SDSS tracers, SDSS tracers allow only a consistency test and not a robust measurement of the kinematic dipole.  

\begin{deluxetable*}{lccccccccc}
\tablecaption{ Solar system proper motion relative to the distant Large Scale Structure (LSS) frame, as derived from observed redshifts obtained through BOSS and eBOSS. The effective redshift ($z_{\rm eff}$) of the LSS tracer is calculated as $\sum_i w_i z_i / \sum_i w_i$, where $w_i$ are weights associated with each object. Error bars and recovery biases are determined using EZmocks for eBOSS and MultiDark-Patchy mocks for BOSS tracers. The lower and upper bounds represent the 16th and 84th percentiles, respectively. Note that the proper motion of the Solar system inferred from the CMB dipole is $369.82 \pm 0.11$ km s$^{-1}$ in the direction  $l=264^\circ.021 \pm 0^\circ.011$, $b=48^\circ.253\pm 0^\circ.005$ (galactic coordinates). \label{tb:results1}}
\centering 
\tablehead{ 
 & LSS tracer & $z_{\rm eff}$  & \multicolumn{3}{c}{CMB dipole as initial guess} & \multicolumn{3}{c}{No initial guess} \\ 
\cmidrule(lr){4-6} \cmidrule(lr){7-9}
            &            &                     & $v$ (km/s) & $l(^\circ)$ & $b(^\circ)$ & 
 $v$ (km/s) & $l(^\circ)$ & $b(^\circ)$
} 
\startdata
\multirow{3}{*}{eBOSS} & QSOs & 1.51  & $377^{+72}_{-74}$ & $266^{+16}_{-16}$ & $49^{+11}_{-11}$  & $134^{+266}_{-134}$ & $262^{+82}_{-85}$ & $-17^{+107}_{-73}$ \\
                       & LRGs & 0.73  & $466^{+97}_{-102}$ & $271^{+23}_{-22}$ & $65^{+11}_{-11}$ & $167^{+202}_{-167}$ & $78^{+59}_{-58}$ & $43^{+47}_{-66}$ \\ 
                       & CMASS & 0.65 & $352^{+54}_{-54}$ & $262^{+11}_{-10}$ & $46^{+9}_{-9}$ & $167^{+193}_{-167}$ & $128^{+122}_{-121}$ & $76^{+14}_{-35}$ \\
                       &&&&&\\
\multirow{2}{*}{BOSS}  & CMASS & 0.52 & $371^{+49}_{-50}$ & $265^{+9}_{-9}$ & $48^{+7}_{-7}$ & $278^{+133}_{-140}$ & $126^{+40}_{-42}$ & $71^{+16}_{-16}$\\
                       & LOWZ  & 0.31 & $107^{+72}_{-74}$ & $301^{+36}_{-38}$ & $14^{+54}_{-58}$ & $128^{+106}_{-107}$ & $296^{+40}_{-41}$ & $-10^{+97}_{-80}$ \\
\enddata
\end{deluxetable*}

We next present the results where we consider the random mocks that include a clustering dipole as described in Sect.~\ref{ssc:error} to correct for recovery bias and error estimations. We have tabulated our results in Table \ref{tb:results2}. It is evident that the inclusion of the clustering dipole does not affect the result significantly, and the random mocks also produce similar results to the EZmocks. We reiterate here that the input clustering dipole we consider in our mocks is \(D = 0.86 \times 10^{-2}\) in the direction \((l, b) = (217^\circ, +20^\circ)\), identified as the residual dipole after subtracting the kinematic dipole in NVSS and WISE samples \citep{Secrest:2022}. However, considering \(\Lambda\)CDM and using Planck $2018$ cosmological parameters \citep{Planck_results:2018} as the fiducial cosmology, along with the respective tracers' \(dN/dz\), we obtain clustering dipole magnitudes of \( (1.6b+1.6) \times 10^{-4}\), \((0.8 b+0.8) \times 10^{-3}\), \((1.5b+1.5) \times 10^{-3}\), \((1.6b + 1.5) \times 10^{-3}\), and \( (3.2b+2.3) \times 10^{-3}\) for tracers QSO, LRG, eBOSS CMASS, BOSS CMASS, and LOWZ, respectively. The first term from the density is multiplied by a galaxy bias factor which typically takes values between 1 -- 3. The second term comes from redshift-space distortions and is assumed to be unbiased.  We obtain these numbers by employing the Core Cosmology Library (CCL; \citealt{Chisari:2019}) and computing the angular auto-spectrum for tracers. The dipole term of the power spectrum, \(C_1\), corresponds to a dipole amplitude \(\mathcal{D}\) as \(C_1 = \frac{4 \pi}{9} \mathcal{D}^2\). Note that this  clustering dipole expected in the context of the $\Lambda$CDM model is smaller than the value $D$ which we insert in our mocks for all SDSS tracers except LOWZ for which it is roughly equal when assuming a bias of $b\simeq 2$.

\begin{deluxetable*}{lccccccccc}
\tablecaption{Solar system proper motion relative to the distant Large Scale Structure (LSS) frame. Error bars and recovery biases are determined using random mocks with and without clustering dipole inclusion. All results are obtained using the CMB dipole as the initial guess; other details remain the same as in Table \ref{tb:results1}. \label{tb:results2} }
\centering 
\tablehead{ 
 & LSS tracer & $z_{\rm eff}$  & \multicolumn{3}{c}{No clustering dipole} & \multicolumn{3}{c}{With clustering dipole} \\ 
\cmidrule(lr){4-6} \cmidrule(lr){7-9}
            &            &                     & $v$ (km/s) & $l(^\circ)$ & $b(^\circ)$ & 
 $v$ (km/s) & $l(^\circ)$ & $b(^\circ)$ \\ 
} 
\startdata
\multirow{3}{*}{eBOSS} &  QSOs & 1.51  & $368^{+92}_{-95}$ & $265^{+16}_{-16}$ & $48^{+14}_{-15}$ & $374^{+94}_{-93}$ & $265^{+17}_{-17}$ & $48^{+14}_{-14}$ \\
                       &  LRGs & 0.73  & $457^{+111}_{-111}$ & $270^{+24}_{-24}$ & $64^{+12}_{-13}$ & $461^{+110}_{-107}$ & $270^{+25}_{-24}$ & $65^{+13}_{-13}$ \\
                       &  CMASS & 0.65 & $348^{+59}_{-56}$ & $262^{+12}_{-12}$ & $45^{+10}_{-9}$ & $349^{+57}_{-60}$ & $263^{+12}_{-12}$ & $45^{+9}_{-10}$ \\  
                       &&&&&\\
\multirow{2}{*}{BOSS}  & CMASS & 0.52 & $371^{+102}_{-96}$ & $264^{+15}_{-14}$ & $48^{+15}_{-14}$ & $373^{+98}_{-102}$ & $264^{+14}_{-14}$ & $48^{+14}_{-15}$ \\
                       &  LOWZ &  0.31 & $95^{+68}_{-71}$ & $302^{+40}_{-40}$ & $10^{+64}_{-65}$ & $97^{+72}_{-73}$ & $302^{+39}_{-40}$ & $12^{+62}_{-67}$ \\
\enddata
\end{deluxetable*}

Finally, we employ bootstrap sampling to derive error bars directly from the data. We generate 1000 bootstrap samples by resampling with replacement from the mock data. For each sample, we calculate the velocity vector components \( v_x \), \( v_y \), and \( v_z \), and then convert the mean and variances of these Cartesian components to spherical coordinates to obtain \( v \), \( l \), and \( b \) along with their error bars. Table \ref{tb:resBootstrap} displays the results for various tracers with and without the CMB dipole as the starting point for the optimization algorithm. The bootstrap sampling provides error estimates robust to sampling variability, offering a reliable measure of uncertainties in our results independent of any fiducial dipole assumption. While using the CMB dipole as the starting point for the optimization algorithm yields results similar to those obtained with EZmocks and random mocks. However, omitting a specific starting point results in estimates with relatively lower values of velocity, also consistent with zero. This observation further supports that the data, at present, serves more as a consistency test rather than a robust measurement of the kinematic dipole.

\begin{deluxetable*}{lccccccccc}
\tablecaption{Solar system proper motion relative to the distant Large Scale Structure (LSS) frame. Error bars are determined using 1000 bootstrap samples. The results are in the format: data (mean from bootstrap samples) $^{+\sigma_{+}}_{-\sigma_{-}}$, without any recovery bias correction. Other details remain the same as in Table  \ref{tb:results1}.\label{tb:resBootstrap}}.
\centering 
\tablehead{ 
 & LSS tracer & $z_{\rm eff}$  & \multicolumn{3}{c}{CMB dipole as initial guess} & \multicolumn{3}{c}{No initial guess} \\ 
\cmidrule(lr){4-6} \cmidrule(lr){7-9}
            &            &                     & $v$ (km/s) & $l(^\circ)$ & $b(^\circ)$ & 
 $v$ (km/s) & $l(^\circ)$ & $b(^\circ)$ \\ 
} 
\startdata
\multirow{3}{*}{eBOSS} & QSOs & 1.51  & $370(351)^{+85}_{-66}$ & $264(265)^{+19}_{-16}$ & $48(54)^{+13}_{-10}$   & $127(51)^{+293}_{-127}$ & $252(146)^{+93}_{-95}$ & $-12(33)^{+102}_{-78}$ \\
                       & LRGs & 0.73  & $462(282)^{+131}_{-130}$ & $270(267)^{+27}_{-24}$ & $65(58)^{+17}_{-15}$ & $162(158)^{+198}_{-162}$ & $78(104)^{+66}_{-64}$ & $45(6)^{+45}_{-72}$ \\
                       & CMASS & 0.65 & $349(329)^{+78}_{-64}$ & $263(267)^{+16}_{-10}$ & $45(54)^{+13}_{-10}$   & $173(5)^{+111}_{-120}$ & $127(99)^{+96}_{-91}$ & $77(68)^{+13}_{-23}$ \\
                       &&&&&\\
\multirow{2}{*}{BOSS}  & CMASS & 0.52 & $370(261)^{+158}_{-192}$ & $264(270)^{+26}_{-25}$ & $48(82)^{+24}_{-30}$ & $280(167)^{+235}_{-235}$ & $129(94)^{+76}_{-68}$ & $71(52)^{+19}_{-26}$ \\
                       & LOWZ  & 0.31 & $98(181)^{+114}_{-98}$ & $302(274)^{+52}_{-46}$ & $11(45)^{+79}_{-98}$   & $126(66)^{+100}_{-105}$ & $296(69)^{+38}_{-41}$ & $-11(-81)^{+76}_{-79}$ \\
\enddata
\end{deluxetable*}

\section{Discussion and Conclusion}
\label{sc:dc}

In this paper we attempted to obtain an independent measure of the kinematic dipole using BOSS and eBOSS data. By employing clustering catalogs, we estimated the best-fit redshift dipole to the data and obtained the Solar systems local motion relative to the distant LSS frame. Our results, summarized in Tables \ref{tb:results1}, \ref{tb:results2} and \ref{tb:resBootstrap}, are consistent with the CMB kinematic dipole for all tracers, except for the LOWZ sample without significant signal (even less than $2\sigma$). This consistency is particularly evident when we use an initial guess close to the CMB-inferred kinematic dipole. The LOWZ tracers are at the lowest redshifts, and somehow the redshift dipole in these galaxies is inconsistent with the CMB dipole. Maybe peculiar velocities and bulk flows play a significant role here, although the redshift for the LOWZ sample is high enough that we should not see such an inconsistency. However, with bootstrap samples, we observe a significant scatter in velocity for the LOWZ sample, similar to what is seen with other tracers. The mean velocity from all bootstrap samples is nearly twice that of the value extracted from the full data. This suggests that the mean from the full data may not accurately represent the population. Nevertheless, the error bars for the recovered velocity and dipole direction for all tracers are quite significant due to limited survey sky coverage and number density. Furthermore, the best dipole value largely depends on the starting point for the optimization algorithm. Changing the starting point or considering no particular initial value of $\vec \beta$ leads to unstable results, which are consistent both with CMB values and with zero.  This suggests that the current SDSS data do not robustly test the kinematic dipole but exhibit consistency with the CMB kinematic dipole when we initiate the optimization algorithm \texttt{scipy.optimize.minimize} with the CMB kinematic dipole itself. Thus, no strong conclusive results on the kinematic dipole can be derived from these SDSS LSS catalogs.
A conservative upper limit at 95 \% C.L. on the velocities can be obtained from the bootstrap resampling. We find the consistent limits of $v < 576$ km/s, $579$ km/s, $514$ km/s, $538$ km/s, and $420$ km/s for the QSO, LRG, eBOSS CMASS, BOSS CMASS, and LOWZ samples, respectively. These upper limits are up to a factor of 1.6 larger than the CMB inferred velocity of the Solar system and cannot explain the excess source count dipoles that are found for radio continuum sources and quasars.

With mocks we have demonstrated that including the clustering dipole  \footnote{Alternatively, one can refer to this as an intrinsic dipole.} does not significantly alter our findings, reinforcing the robustness of the redshift dipole in recovering the kinematic dipole. A plausible reason is that the redshift dipole arises from the radial shift of galaxy positions and is largely independent of galaxy clustering due to LSS. This suggests that the measurement using the redshift dipole captures primarily the kinematic component, and consistency with the CMB kinematic dipole does not imply isotropy of the matter distribution or adherence to the cosmological principle. LSS surveys such as NVSS and WISE exhibit a relatively large dipole, suggesting a violation of large-scale isotropy and homogeneity, fundamental to the cosmological principle. It is noteworthy that the NVSS and WISE dipoles are the vector sum of kinematic, clustering, and shot-noise components. Independently, this work shows consistency of the kinematic dipole with the CMB, indicating that the significant dipoles observed in NVSS and WISE-like catalogs are likely attributed primarily to the clustering dipole \footnote{By ``clustering dipole," we refer to the total dipole arising from galaxy clustering, not just the dipole expected from $\Lambda$CDM.  It is noteworthy that, in addition to the small clustering dipole predicted by $\Lambda$CDM (as presented in Section \ref{sc:results}), several alternative explanations have been proposed to account for the observed excess dipole through galaxy clustering effects. These include the presence of rare voids or superclusters of varying scales (e.g., \citealt{Rubart:2014, Cai:2022}), superhorizon-sized density fluctuations \citep{Tiwari:2022SP, Domenech:2022}, and an unexpectedly large and evolving clustering bias of AGNs \citep{Adi:2015nb, Tiwari:2016adi}.} ,if not due to some observational or analysis systematics.

Despite the small recovery bias observed, our results indicate that the current SDSS data serves as a consistency test rather than a definitive measure of the kinematic dipole. The limited sky coverage of the SDSS tracers introduces inconsistency and instability in minimization results, which we have mitigated by using the CMB kinematic dipole as the initial guess in our minimization process. Future surveys with more extensive sky coverage and larger density of spectroscopic redshifts will be crucial for achieving a more robust test. Currently, it suggests that the kinematic dipole appears consistent with the CMB, implying that any observed dipolar anisotropy in number density or flux with LSS tracers is likely due to clustering. This excess clustering could be attributed to systematics, a relatively large number of local sources, or biases and uncertainties in flux measurements. If this is not the case, then the observations from NVSS and WISE stand as a potential violation of the cosmological principle.

\begin{acknowledgments} 
G.B.Z. is supported by National Key R\&D Program of China No. 2023YFA1607803, NSFC grant 11925303, and by the CAS Project for Young Scientists in Basic Research (No. YSBR-092), the China Manned Space Project with No. CMS-CSST-2021-B01, and the New Cornerstone Science Foundation through the XPLORER prize. D.J.S. acknowledges financial support by the ``NRW-cluster for data intensive radio astronomy: Big Bang to Big Data (B3D)'' funded through the programme ``Profilbildung 2020'', an
initiative of the Ministry of Culture and Science of the State of North Rhine-Westphalia. R.D. and M.K. acknowledge funding by the Swiss National Science Foundation. H.P. acknowledges support from the Swiss National Science Foundation via Ambizione Grant PZ00P2\_179934.

\end{acknowledgments}

\appendix

\section{Data Availability}
\label{sc:datalinks}
The eBOSS clustering catalogs can be accessed at \url{https://data.sdss.org/sas/dr17/eboss/lss/catalogs/DR16/}. We utilized the following files: \\ \texttt{eBOSS\_*\_clustering\_data-NGC/SGC-vDR16.fits}, where * represents tracers such as QSO, LRG, and LRGpCMASS.\\
The EZmocks are available at \url{https://data.sdss.org/sas/dr17/eboss/lss/EZmocks/v1_0_0/realistic/eBOSS_*/dat}, where * again denotes tracers.
\\
BOSS clustering catalogs can be found at \url{https://data.sdss.org/sas/dr12/boss/lss/}. Specifically, we used files  \texttt{galaxy\_DR12v5\_*\_North/South.fits.gz}, where * denotes CMASS or LOWZ tracers. \\
The MultiDark Patchy Mocks for BOSS are accessible at \url{https://data.sdss.org/sas/dr12/boss/lss/dr12_multidark_patchy_mocks/}. We used \texttt{Patchy-Mocks-DR12NGC/SGC-COMPSAM\_V6C.tar.gz} in our analysis.

\software{HEALPix \citep{Gorski:2005}, Core Cosmology Library \citep{Chisari:2019}, CAMB \citep{Challinor:2011}. }

\bibliographystyle{apj}
\bibliography{master}
\end{document}